\documentclass{jfm}
\usepackage{graphicx}
\usepackage{epstopdf, epsfig}
\usepackage[ruled,vlined]{algorithm2e}
\usepackage{xcolor}
\usepackage{xr}

\usepackage{pdfpages}

\makeatletter
\newcommand*{\addFileDependency}[1]{
  \typeout{(#1)}
  \@addtofilelist{#1}
  \IfFileExists{#1}{}{\typeout{No file #1.}}
}
\makeatother

\newcommand*{\myexternaldocument}[1]{%
    \externaldocument{#1}%
    \addFileDependency{#1.tex}%
    \addFileDependency{#1.aux}%
}

\myexternaldocument{supplementaryMaterials}

\newtheorem{remark}{Remark}

\shorttitle{Inferring 3D flow over an espresso cup based on Tomo-BOS/PINN}
\shortauthor{S. Cai, Z. Wang, F. Fuest, Y. Jeon, C. Gray and G.E. Karniadakis}

\title{Flow over an espresso cup: Inferring 3D velocity and pressure fields from tomographic background oriented {schlieren} videos via physics-informed neural networks}

\author{Shengze Cai\aff{1},
  Zhicheng Wang\aff{1},
  Frederik Fuest\aff{2},
  Young Jin Jeon\aff{2},
  Callum Gray\aff{3}
 \and George Em Karniadakis\aff{1}
 \corresp{\email{george\_karniadakis@brown.edu}}
 }

\affiliation{\aff{1}Division of Applied Mathematics, Brown University, Providence, RI, 02912, USA
\aff{2}LaVision GmbH, Anna-Vandenhoeck-Ring 19, D-37081 Goettingen, Germany
\aff{3}LaVision Inc., 211 W. Michigan Ave., Ypsilanti, MI 48197, USA
}

\begin{document}

\maketitle

\begin{abstract}
Tomographic background oriented schlieren (Tomo-BOS) imaging measures density or temperature fields in 3D using multiple camera BOS projections, and is particularly useful for instantaneous flow visualizations of complex fluid dynamics problems.
We propose a new method based on physics-informed neural networks (PINNs) to infer the full continuous 3D velocity and pressure fields from snapshots of 3D temperature fields obtained by Tomo-BOS imaging. 
PINNs seamlessly integrate the underlying physics of the observed fluid flow and the visualization data, hence enabling the inference of latent quantities using limited experimental data. In this hidden fluid mechanics paradigm, we train the neural network by minimizing a loss function composed of a data mismatch term and residual terms associated with the coupled Navier-Stokes and heat transfer equations.
We first quantify the accuracy of the proposed method based on a 2D synthetic data set for buoyancy-driven flow, and subsequently apply it to the Tomo-BOS data set, where we are able to infer the instantaneous velocity and pressure fields of the flow over an espresso cup based only on the temperature field provided by the Tomo-BOS imaging. Moreover, we conduct an independent PIV experiment to validate the PINN inference for the unsteady velocity field at a center plane. To explain the observed flow physics, we also perform systematic PINN simulations at different Reynolds and Richardson numbers and quantify the variations in velocity and pressure fields. The results in this paper indicate that the proposed deep learning technique can become a promising direction in experimental fluid mechanics.

\end{abstract}


\section{Introduction } \label{sec:intro}

Background oriented schlieren (BOS) imaging has become an effective technique for flow visualization and quantitative fluid measurement in recent years \citep{raffel2000applicability,richard2001principle,raffel2015background}. Compared to other flow visualization technologies, such as particle image velocimetry (PIV) \citep{raffel2018particle} and laser induced fluorescence (LIF) \citep{crimaldi2008planar}, BOS is less expensive and more flexible to set up in different experimental environments. Schlieren photography is generally used to visualize the flow that contains density gradients, which are caused by varying temperature, pressure or composition in species. 
In principle, BOS is based on the fact that an object inside a varying density medium will appear distorted due to the refraction of light rays. The distortion is then evaluated by using digital image correlation methods \citep{pan2018digital} for the reference image (without density gradients) and the distorted images. Alternatively, optical fow algorithms \citep{atcheson2009evaluation} and dot tracking methods \citep{rajendran2019dot} can also be applied to reconstruct the displacement field characterizing the distortion. 
In addition to the refractive index and the density, temperature field or other quantities can be further inferred by analyzing the BOS images \citep{tokgoz2012temperature}.
In the past two decades, the BOS technology has been applied to various applications in fluid mechanics \citep{venkatakrishnan2004density,goldhahn2007background,grauer2018instantaneous,nicolas2017experimental}. However, despite the substantial improvements of the hardware (e.g., from two-dimensional setup to tomographic setup \citep{raffel2000applicability}), quantifying the entire fluid velocity and pressure fields continuously in space-time from BOS images has remained an open problem.

The existing algorithms for BOS velocimetry are mostly developed based on the cross-correlation method, which determines the displacement between two interrogation windows by searching the maximum of the correlation. 
For example, \citet{buhlmann2014laser} performed PIV analysis on two consecutive BOS speckle displacement fields and obtained a spatially resolved estimate of local convection velocities. 
However, the correlation-based methods are mostly optimized for PIV technology (which requires tracer particles in the flow) rather than BOS images (which usually visualizes a scalar field). 
\citet{raffel2011density} proposed a method called density tagging velocimetry to overcome this problem, in which the local density variation acts as a tracer particle transported by a fluid flow. 
%
These existing algorithms generally produce sparse velocity vectors as the computations are based on local patterns of the BOS images. Moreover, these methods are not able to estimate the pressure from BOS data directly. A pressure reconstruction method, such as solving the Poisson equation, is required for pressure inference, which is used routinely in PIV \citep{wang2016irrotation,zhang2020using}. 
In addition, reconstructing the three-dimensional (3D) velocity field from tomographic BOS (Tomo-BOS) is still an open area in experimental fluid mechanics.

In this paper, we develop a new method for estimating the continuous velocity and pressure fields simultaneously from Tomo-BOS data. Here, the Tomo-BOS data that we are interested are the 3D temperature fields at different time instants.  Inspired by the development of deep learning methods \citep{lecun2015deep}, we exploit the expressivity of deep neural networks to address this problem.
Recently, deep learning techniques have become increasingly popular in both computational and experimental fluid dynamics \citep{brunton2020machine,duraisamy2019turbulence}.
For example, \citet{raissi2019physics} proposed a framework of physics-informed neural networks (PINNs) to solve forward and inverse problems of nonlinear partial differential equations (PDEs), \citet{han2018solving} and \citet{sirignano2018dgm} also proposed to use deep learning methods to handle high-dimensional PDEs. These recent advances of scientific machine learning have also inspired innovation in fluid mechanics.
The PINN framework \citep{raissi2019physics} has also been applied to perform complex flow simulations \citep{sun2020surrogate,jin2020nsfnets}. On the other hand, data-driven algorithms have  also been investigated for dealing with real experimental data \citep{rabault2017performing,cai2019particle,jin2020time}. 
For instance, a convolutional neural network (CNN) was applied to infer a dense velocity field from PIV images by \citet{cai2019particle}. Moreover, \citet{raissi2020hidden} proposed the ``hidden fluid mechanics" (HFM) method based on PINNs to integrate the Navier-Stokes equations and the visualization data, which enables quantification of the velocity and pressure from 3D concentration fields.
HFM, which so far has been applied only to synthetic data, is similar to data assimilation, however, the solutions of the continuous flow fields are approximated by a neural network rather than solved
by computational fluid dynamics methods.

\begin{figure}
\begin{center}
\includegraphics[width=\textwidth]{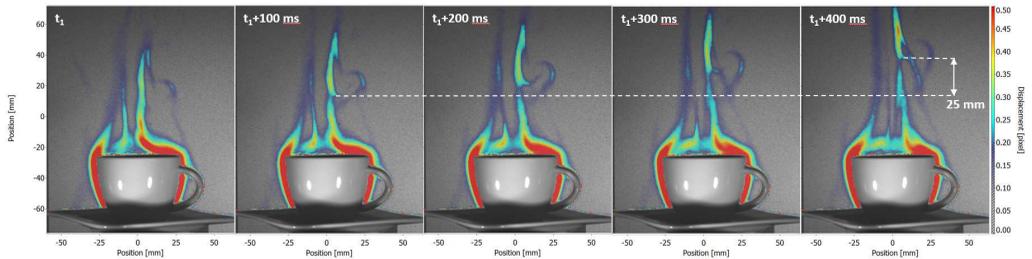}
\caption{Illustration of a sequence of temperature-induced schlieren images from one camera observing the flow developing over an espresso cup. A video of the schlieren images is provided in the supplementary material.
}
\label{fig:intro_2D_image}
\end{center}
\end{figure}

Our work is inspired by \citet{raissi2020hidden} as we focus on the velocity and pressure inference of  buoyancy-driven flows when the temperature is determined by Tomo-BOS imaging.
We present a Tomo-BOS experiment to observe the flow that takes place over an espresso cup. A sequence of five schlieren images from one of the cameras in the experiment is illustrated in Figure \ref{fig:intro_2D_image}.
We propose to use physics-informed neural network, which is capable of seamlessly integrating the governing equations of the natural convection problem and the reconstructed 3D temperature data. 
To our knowledge, this is the first time that the HFM paradigm has been applied to the real experimental imaging data. 
%

The rest of this paper is organized as follows. We introduce the methodology combining neural network and the physical  model in Sect. \ref{sec:PINNs}. 
The results of applying PINN algorithm to Tomo-BOS imaging are presented in Sect. \ref{sec:TomoBOS_exp}, where we quantify the 3D velocity and pressure of the flow over an espresso cup based on the imaged temperature field. We also conduct a companion PIV experiment to obtain the unsteady velocity field at the center plane for validation purposes. 
Finally, we present
the conclusion in Sect. \ref{sec:conclusion}. 
Note that in the supplementary material (SM), we also evaluate the proposed method based on synthetic data (i.e., 2D simulated flow), and investigate the performance of the method with respect to different parameter settings, including the hyper-parameters of the neural network, the data resolution and the signal-to-noise ratio. 

\section{Physics-informed neural networks (PINNs)}
\label{sec:PINNs}

The physics-informed neural networks (PINNs), which is also called ``hidden fluid mechanics" (HFM), can infer the latent quantities such as the velocity and pressure fields from auxiliary data, e.g., from smoke
visualizations. In this work, we extend PINNs to deal with visualized temperature data of buoyancy-driven flow from Tomo-BOS imaging experiments.
We employ the Boussinesq approximation of the incompressible Navier-Stokes (NS) equations and the corresponding heat transfer equation into PINNs.

Let $T(\mathbf{x},t)$, $\mathbf{u}(\mathbf{x},t)$ and $p(\mathbf{x},t)$ denote the temperature, velocity vector and pressure, respectively, where $\mathbf{x}\in\mathbb{R}^2$ or $\mathbb{R}^3$ represents the spatial coordinate and $t$ is the temporal coordinate.
The main idea of PINNs is to approximate the spatio-temporal solutions by using a fully-connected neural network (FNN). This can be represented as:
\begin{equation}
(T, \mathbf{u}, p) = \mathcal{F}_{NN}(\mathbf{x},t, \Theta),
\label{eq:PINN_approximation}
\end{equation}
where $(\mathbf{x},t)$ and $(T, \mathbf{u}, p)$ denote the inputs and outputs of the neural network, respectively, and $\Theta$ denotes the trainable parameters. For the fully-connected network, $\Theta$ includes the weights and biases of multiple hidden layers. For the $k$-th hidden layer, the relation between the output vector $Y^{k}$ and the input vector $X^{k}$ can be simply expressed as:
\begin{equation}
Y^{k} = \sigma (W^{k}X^{k}+b^{k}),
\label{eq:hidden_layer}
\end{equation}
where $W^{k}$ and $b^{k}$ are the weights and biases, and $\sigma(\cdot)$ denotes the activation function, which is used to represent the nonlinearity of the solutions. We use the hyperbolic tangent function, namely $\sigma(\cdot)=\tanh(\cdot)$, throughout the paper unless otherwise stated.
In this context, solving the nonlinear system of the governing equations is equivalent to learning the weights and biases of the network. However, this is an extremely ill-posed problem as we assume that only the temperature data $T$ is given without any boundary conditions on the velocity field. In order to infer the velocity and pressure fields,
an additional network with the underlying physics is encoded into PINNs. Specifically, we aim to minimize the residuals of all the governing equations  in the additional networks, which can be thought of as enforcing strong constraints for the feed-forward neural network outputs.
For $\mathbf{x}\in\mathbb{R}^3$, the residuals are defined as follows:
\begin{subeqnarray}
e_{1} &=& T_{t} + uT_{x}+vT_{y}+wT_{z} - {1}/{\mbox{Pe}} (T_{xx}+T_{yy}+T_{zz}) ,  \\
e_{2} &=& u_{t} + uu_{x}+vu_{y}+wu_{z} + p_{x} -  {1}/{\mbox{Re}}(u_{xx}+u_{yy}+u_{zz}) + \mbox{Ri}T\mathbf{e_{x}}  ,  \\
e_{3} &=& v_{t} + uv_{x}+vv_{y}+wv_{z} + p_{y} -  {1}/{\mbox{Re}}(v_{xx}+v_{yy}+v_{zz}) + \mbox{Ri}T\mathbf{e_{y}}   ,  \\
e_{4} &=& w_{t} + uw_{x}+vw_{y}+ww_{z} + p_{z} -  {1}/{\mbox{Re}}(w_{xx}+w_{yy}+w_{zz}) + \mbox{Ri}T\mathbf{e_{z}}   ,  \\ 
e_{5} &=& u_{x}+v_{y}+w_{z},
\label{eq:PINN_residuals_3D}
\end{subeqnarray}
where the subscripts represent the derivatives of the corresponding quantities, $(\mathbf{e_{x}}, \mathbf{e_{y}}, \mathbf{e_{z}})$ are the components of the gravity unit $\mathbf{e_{g}}$ in $(x,y,z)$ directions. 
Here, the residuals $e_{1}-e_{5}$ correspond to the dimensionless heat transfer equation and the Boussinesq approximation of incompressible NS equations. For the buoyancy-driven flow, it is assumed that the density variation consists of a fixed part and another part that has a linear dependence on temperature, where the latter leads to the gravity force term in  Eq.~(\ref{eq:PINN_residuals_3D}b)-(\ref{eq:PINN_residuals_3D}d).
We also note that the pressure featured in the momentum equations must be considered as the pressure correction representing the deviation from hydrostatic equilibrium, which is not constant and not negligible when solving the equations. 
The non-dimensional parameters, including Reynolds number, P\'{e}clet number and Richardson number, are defined as follows:
\begin{equation}
\mbox{Re} = \frac{UL}{\nu}, \quad \mbox{Pe} = \frac{UL}{\alpha}, \quad  \mbox{Ri} = \frac{g\beta(T_{hot}-T_{\infty})L}{U^2},
\label{eq:non_dimensional_para}
\end{equation}
where these physical properties will be specified in the experiment.
%

\begin{figure}
\begin{center}
\includegraphics[width=\textwidth]{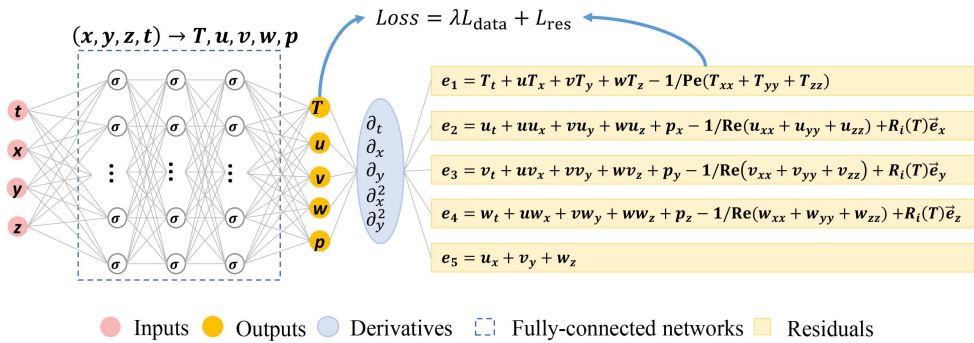}
\caption{Schematic diagram of the physics-informed neural network for the 3D velocity and pressure estimation. The neural network is composed of a fully-connected network and a residual network. The fully-connected network is used to represent the solutions of the velocity, pressure and temperature fields, while the residual network is considered as the physical constraints (i.e., the incompressible Navier-Stokes equations and the heat equation) for the solutions. 
The parameters of the network are learned by minimizing a loss function including a data term and the residuals of the equations.   }
\label{fig:PINNs}
\end{center}
\end{figure}

A schematic illustration of the proposed physics-informed neural network, composed of a FNN with multiple hidden layers and a residual network with the physical constraints, is shown in Figure \ref{fig:PINNs}. 
In order to compute the residuals $e_{1}-e_{5}$, the derivatives of the state variables with respect to space and time are required, which is achieved by using automatic differentiation (AD) in the deep learning code \citep{baydin2017automatic}.
AD relies on the chain rule of derivative computation and it is well-implemented in most deep learning frameworks such as TensorFlow and PyTorch, that allows us to avoid numerical discretization while computing derivatives of all orders in space-time. Once the FNN and the residuals have been formulated, the parameters $\Theta$ of PINNs are trained by minimizing the following loss function:
\begin{equation}
\mathop{\arg\min}_{\Theta} \quad L = \lambda L_{\mathrm{data}} +  L_{\mathrm{res}},
\label{eq:loss}
\end{equation}
where $\lambda$ is a weighting coefficient and
\refstepcounter{equation}
$$
L_{\mathrm{data}} = \sum_{n=1}^{N_T} |T(\mathbf{x}^{n},t^{n}) - T^{n}_{\mathrm{data}} |^{2},  \quad
L_{\mathrm{res}} = \sum_{i} \sum_{n=1}^{N_{e}} |e_{i}(\mathbf{x}^{n},t^{n}) |^{2}.
$$
Here, $L_{\mathrm{data}}$ represents the mismatch between the temperature data and the temperature predicted by the neural network, $L_{\mathrm{res}}$ represents the residuals, and $N_{T}$ and $N_{e}$ are the numbers of training points corresponding to two terms. Note that the value of $N_{T}$ depends on the grid points (e.g., pixel or voxel) of the observed data, while the value of $N_{e}$ can be very large since the spatio-temporal points for residual network can be randomly selected in the domain and the mini-batch technique can be employed.
The weighting coefficient $\lambda$ is applied to adjust the contributions of two terms in the loss function and we choose $\lambda=100$ for the Tomo-BOS experiment in this paper. We note that a large weight of the data term $L_{\mathrm{data}}$ can accelerate the convergence of training, but it may weaken the constraint of the governing equations and result in overfitting. The sensitivity of the hyperparameters in PINN is quantitatively analyzed on a synthetic dataset and is presented in SM. 
By default, the parameters of the neural networks $\Theta$ are initialized using the Xavier scheme, and then optimized until convergence via an adaptive optimization algorithm \emph{Adam} \citep{kingma2014adam}. 
When the minimizer of the loss function (\ref{eq:loss}) is obtained after training, the velocity and pressure fields can be inferred simultaneously with the trained parameters by feeding the spatio-temporal coordinates to the neural network. 
The proposed method is summarized in the Algorithm \ref{algo:PINNs}, where we would like to emphasize that all the information used in the neural network is the temperature data along with the formulation of the governing equations.

\begin{remark}
The PINN algorithm, which can be considered as a novel data assimilation method, estimates the velocity and pressure fields by regressing the visualization data in a spatio-temporal domain. It is known that the efficiency of conventional data assimilation strategies is sensitive to the choice of initial guess of the initial and boundary conditions of velocity and pressure, which are the control variables of these methods to be optimized. 
However, we should note that the trainable variables of the proposed PINN algorithm are the neural network parameters (need to be initialized as well), which are totally different from the conventional methods. Therefore, it is not necessary to provide the initial and boundary conditions of either velocity or pressure. 
\end{remark}

\begin{algorithm}
\SetAlgoLined
\KwData{Temperature data  $\{\mathbf{x}^{n},t^{n},T^{n}\}_{n=1}^{N^{T}}$}{}
\KwResult{Velocity, pressure and temperature  $\{\mathbf{u}(\mathbf{x},t),p(\mathbf{x},t),T(\mathbf{x},t)\}$, where $(\mathbf{x},t)$ can be any points in the computational domain }
 \textbf{Step 1:} Specify the training sets   \\
 ~~~ Training data: $\{\mathbf{x}^{n},t^{n},T^{n}\}_{n=1}^{N^{T}}$ \\ 
 ~~~ Residual training points: $\{\mathbf{x}^{n},t^{n}\}_{n=1}^{N^{e}}$     
 \\
 \textbf{Step 2:} Construct the fully-connected network $\mathcal{F}_{NN}$ with random initialization of network parameters $\Theta$.   
 \\
 \textbf{Step 3:} Formulate the residuals by substituting the outputs of $\mathcal{F}_{NN}$ into the governing equations using AD and other arithmetic operations. \\
 \textbf{Step 4:} Define the loss function and find the best parameters $\Theta^{*}$ by using the Adam optimization method.
 \\
 \textbf{Step 5:} Estimate the velocity, pressure and temperature fields $\{\mathbf{u}(\mathbf{x},t),p(\mathbf{x},t),T(\mathbf{x},t)\}$ by feeding the coordinates $(\mathbf{x},t)$ to the trained network. 
 \\  
 \caption{Algorithm for the Physics-informed neural networks (PINN) for velocity and pressure inference from temperature data.}
 \label{algo:PINNs}
\end{algorithm}
 

\section{Inference of Tomo-BOS Experiment} \label{sec:TomoBOS_exp}

\subsection{Experimental setup of Tomo-BOS }\label{sec:TomoBOS_exp_setup}


\begin{figure}
\includegraphics[width=\textwidth]{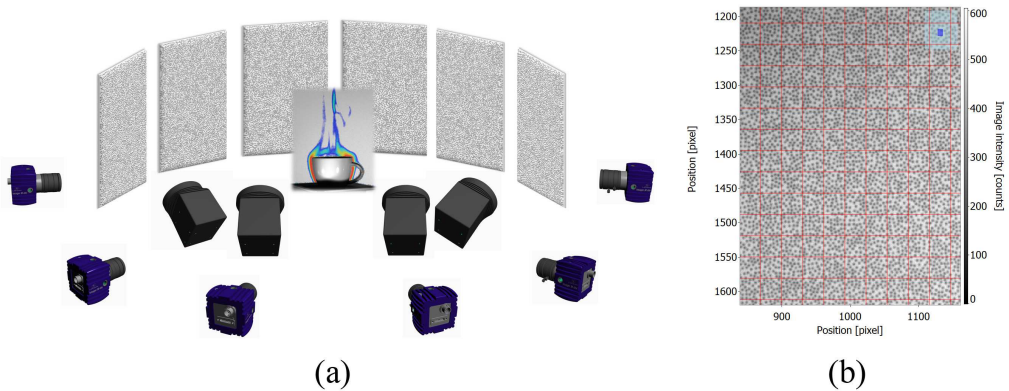}
\caption{Tomo-BOS experiment: (a) Experimental setup for Tomographic BOS measurements around an espresso cup. The espresso cup is shown as an image from one of the cameras with the schlieren image after processing superimposed. (b) Zoomed section of a raw image showing the dots (gray) and subsets (red), where each dot occupies roughly 5 pixels and each subset contains about 25-30 dots. 
}
\label{fig:3D_TomoBOSSetup}
\end{figure}

The experimental setup that was used for acquiring BOS images is illustrated in Figure \ref{fig:3D_TomoBOSSetup}(a). Six cameras (LaVision M-lite 5M, $2056\times2464$ pixel) equipped with 35-mm lenses (f/16) were aligned around an espresso cup equally spaced covering a total angle of 150$^{\circ}$ at a distance of 700~mm from the center of the cup. The projected pixel resolution was 60~$\mu$m/pixel.
Six aluminum sandwich panels ($300 \times 420$ mm$^2$) with imprinted random dot-patterns were placed in the background at a distance of 500~mm from the center of the cup. Four pulsed LEDs (LaVision High Energy LED Spotlight) were used in combination with an exposure time of 500~$\mu$s at a recording rate of 50~Hz to acquire 400 images.

The temperature in the laboratory was 20$^{\circ}$C. 
When the espresso cup was filled with boiling water, 
a thermocouple was used to measure the temperature of the water (2~cm into the water at the center of the cup) and the gas temperature (5~mm above the liquid surface at the center of espresso cup), which are approximately 80$^{\circ}$C and 50$^{\circ}$C, respectively. The estimated uncertainty of these quantities is approximately $\pm3^{\circ}$C. The thermocouple measurements happened about 20~s before the Tomo-BOS data acquisition, while the whole experiment (including the thermocouple measurements and BOS image recording) was performed within one minute.

The image data were acquired and processed using LaVision's Tomographic BOS software (DaVis 10.1.1). As shown in the zoomed image in Figure \ref{fig:3D_TomoBOSSetup}(b), each of the subsets contains about 25-30 dots in a window with 31$\times$31~pixel$^2$, where each dot occupies roughly 5~pixels. In order to determine the 2D displacement fields, a subset-based zero-mean normalized sum of squared difference (ZNSSD) algorithm with first order shape functions \citep{pan2018digital} was employed, 
resulting in a spatial resolution of 600~$\mu$m/pixel. 
Outliers were not observed in the displacement fields and therefore we did not apply any post-processing filters. 

%

An example of the resulting schlieren images is shown in Figure~\ref{fig:intro_2D_image}, which demonstrates the temperature-induced schlieren visualized as colormap of the 2D displacement field in pixel from one camera. 
In order to reconstruct the 3D field, a calibration procedure was performed, where a spatial calibration plate was additionally recorded at different locations and at different rotation angles. 
Eventually, the 2D displacement fields from all six cameras were subjected to the tomographic reconstruction algorithm proposed by \cite{nicolas2016direct}, which yielded a unique and well-behaved solution with the aids of the Tikhonov smoothness regularization and the 3D mask. The resolved volume of the 3D data is approximately $[-64,62]\times[-90, 85]\times[-75,52]$~mm$^3$ (width$\times$height$\times$depth). 
A 3D refractive index field was obtained, having 0.9~Mvoxel on a Cartesian grid with a grid spacing of 1.5~mm. The temperature at each voxel was then calculated by neglecting pressure variation and using $T=T_0 \rho_0 G/(n-1)$, where $G$ is the Gladstone-Dale constant of air ($G=0.2268$~cm$^3$/g), $T_0$ and $\rho_0$ are ambient temperature and density ($T_0=293.15$~K and $\rho_0=1.204$~kg/m$^3$), respectively. 
We note that multiple tests were performed with the Tomo-BOS experiment. While the transient components of the flow were different between different tests (as the flow is unsteady), the core features (e.g., temperature values) remained the same. 
In the following, the PINN method proposed in Sect. \ref{sec:PINNs} is applied to infer the temporal evolution of 3D velocity and pressure from the temperature data of one of the Tomo-BOS experiments. 
For clarity, these experimental results are denoted by Tomo-BOS/PINN hereafter. 
%


\subsection{Inference Results}\label{sec:TomoBOS_exp_results}

\subsubsection{Results of Tomo-BOS/PINN}\label{sec:TomoBOS_PINN}

\begin{figure}
\begin{center}
\includegraphics[width=0.95\textwidth]{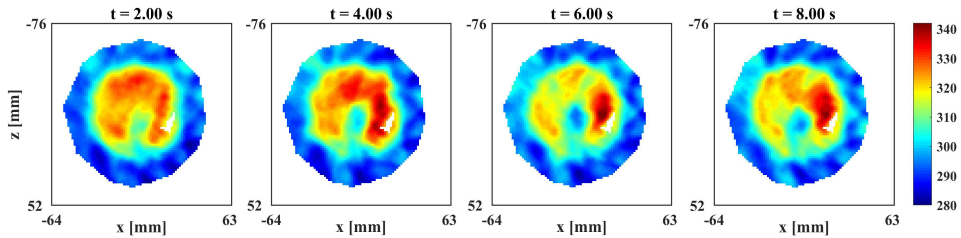}
\caption{Tomo-BOS/PINN results: 2D temperature profiles of Tomo-BOS data on the surface of the espresso cup ($y=-31$~mm) at various time steps. (unit: K)
}
\label{fig:Tomo_2Dhor_temp}
\end{center}
\end{figure}

We first assume that the governing equations of the problem include the
Boussinesq approximation of NS equations and the heat equation, as shown in Eq.~(\ref{eq:PINN_residuals_3D}), where the physical properties appearing in Eq.~(\ref{eq:non_dimensional_para}) are: the thermal diffusivity $\alpha = 2.074 \times 10^{-5} ~ \mbox{m$^{2}$/s}$, the coefficient of thermal expansion $\beta = 3.4 \times 10^{-3} ~ \mbox{1/K}$, the kinematic viscosity of the flow $\nu = 1.516 \times 10^{-5} ~ \mbox{m$^{2}$/s}$ and the gravity acceleration $g = 9.8 ~ \mbox{m/s$^{2}$} $.
In this experiment, we define the characteristic length (which approximates the diameter of the espresso cup) and the characteristic velocity (which is an arbitrary velocity magnitude) as:
$L = 0.065 ~ \mbox{m}$, $U = 0.1 ~ \mbox{m/s}$. 
The environmental temperature is a constant $T_{\infty}=293.15$~K, while determining the reference hot temperature $T_{hot}$ is not straightforward due to the unsteady and non-uniform distribution of the surface temperature above the espresso cup, as illustrated in Figure \ref{fig:Tomo_2Dhor_temp}. Herein, we use the time-averaged value of the maximum surface temperatures from all snapshots, namely $$T_{hot}=1/400\sum_{k=1}^{400}\max(T^{k}),$$ 
which is $ T_{hot} = 338.2$~K.
Then, the non-dimensional parameters, namely the Reynolds number, P\'{e}clet number and Richardson number, can be computed: 
$\mbox{Re} \approx 430$, $\mbox{Pe} \approx 313$, $\mbox{Ri} \approx 9.9$. In the PINN algorithm, we use a neural network with 10 hidden layers and 150 neurons per layer. The weighting coefficient in the loss function is 100. We train the neural network with 120 epochs until the loss reaches a plateau, which takes about 10 hours on a single GPU (for processing 400 snapshots in 3D).

\begin{figure}
\begin{center}
\includegraphics[width=0.95\textwidth]{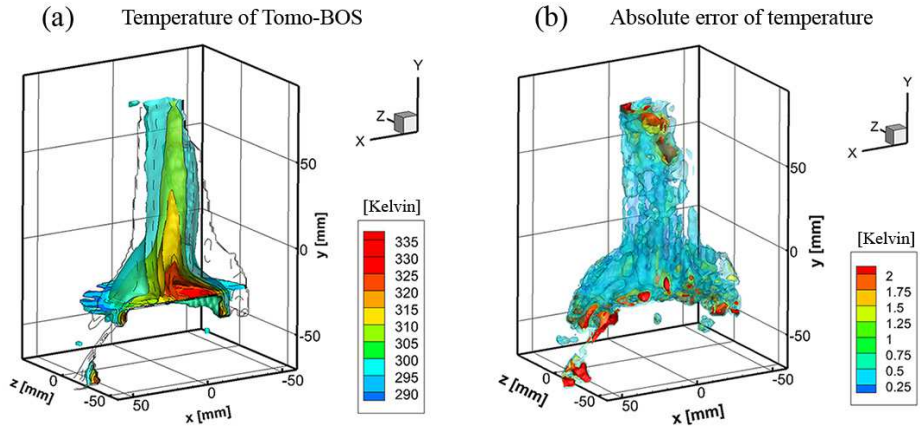}
\caption{Tomo-BOS/PINN results: (a) iso-surfaces of temperature field (corresponding to values of 300~K, 310~K, 315~K and 325~K) at $t=2.0$ s extracted from Tomo-BOS data, (b) iso-surfaces of the difference between Tomo-BOS data and the temperature regressed by PINN, namely $|\hat{T}_{PINN} - T_{BOS}|$. The $L_2$ error of temperature for PINN is less than 1\%.
}
\label{fig:Tomo_temperature}
\end{center}
\end{figure}

In a real experiment, we cannot directly compute the errors of the reconstructed velocity and pressure fields since the ground-truth fields are not available. Therefore, we first examine the regression error of the temperature data, which is an indicator of the convergence of the neural network. 
The iso-surfaces of the 3D temperature fields at $t= 2.0$ s are illustrated in Figure \ref{fig:Tomo_temperature}. The Tomo-BOS data is shown in Figure \ref{fig:Tomo_temperature}(a), while the difference between the data and the temperature regressed by PINN is shown in Figure \ref{fig:Tomo_temperature}(b). 
From the plots we can observe that the temperature field regressed by PINN is almost identical with the Tomo-BOS data, as the absolute error of temperature over the whole spatial domain is less than 1$^{\circ}$C; the corresponding relative $L_2$-norm error (defined by $\epsilon_{T}= \parallel \hat{T}_{PINN} - T_{BOS} \parallel_{2} / \parallel T_{BOS}\parallel_{2}$) is less than 1\%.

\begin{figure}
\begin{center}
\includegraphics[width=0.95\textwidth]{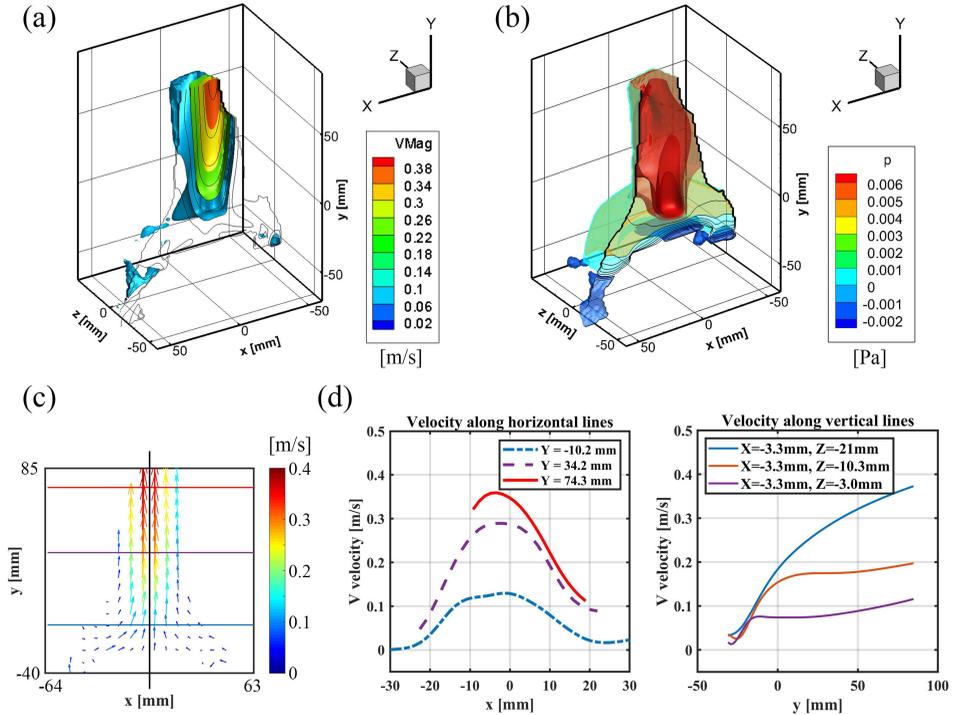}
\caption{Tomo-BOS/PINN results: (a) 3D contours of velocity magnitude, (b) 3D contours of pressure field, (c) 2D velocity vectors at a vertical plane $z=-21$~mm, (d) velocity profiles along three horizontal lines and three vertical lines, which are shown in (c). The results demonstrated here are extracted at time instant $t=2.0$~s. The center of the coffee cup is approximately located at $x=-3.3$ mm and $z=-21$ mm. 
}
\label{fig:TomoBOS_vel}
\end{center}
\end{figure}

\begin{figure}
\begin{center}
\includegraphics[width=0.95\textwidth]{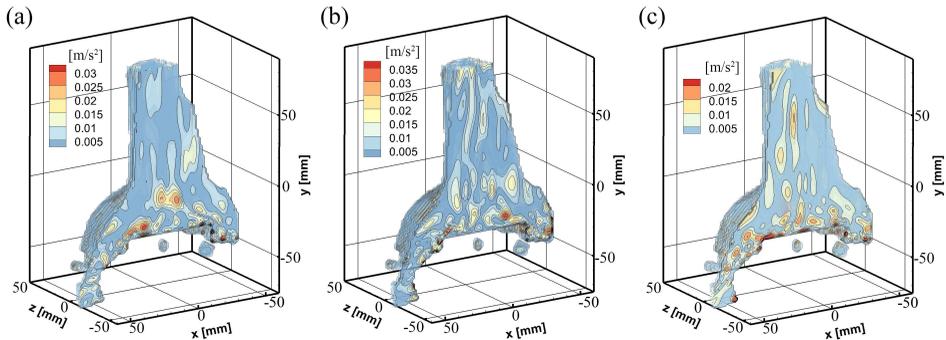}
\caption{Tomo-BOS/PINN results: residuals of the momentum equations in (a) x-direction (Eq.~\ref{eq:PINN_residuals_3D}b), (b) y-direction (Eq.~\ref{eq:PINN_residuals_3D}c), (c) z-direction (Eq.~\ref{eq:PINN_residuals_3D}d). Snapshot at $t=2.0$~s is demonstrated. The average residuals over the spatial domain are
in the order of $10^{-4}$~m/s$^{2}$.
}
\label{fig:Tomo_residual}
\end{center}
\end{figure}

The iso-surfaces of the inferred 3D velocity field at $t=2.0$~s are presented in Figure \ref{fig:TomoBOS_vel}(a), corresponding to velocity magnitudes of 0.1 m/s, 0.15 m/s, 0.25 m/s and 0.35 m/s. 
It can be seen that the velocity contours present an inverted cone shape and the flow speed above the espresso cup increases with height. 
Moreover, the contours of 3D pressure field at $t=2.0$~s are illustrated in Figure \ref{fig:TomoBOS_vel}(b). As mentioned in Sect. \ref{sec:PINNs}, the pressure inferred by PINN is the one featured in the momentum equations, which should be considered as the deviation from hydrostatic equilibrium. As shown, the lowest pressure can be observed on the surface of the espresso cup, where the temperature is high and the 
air flow speed is nearly zero. Then, the pressure increases suddenly above the surface and decreases slightly toward the environment. 
Furthermore, the 2D velocity vectors (at $z=-21$ mm and $t=2.0$ s) and the corresponding 1D velocity profiles along various lines are presented in Figure \ref{fig:TomoBOS_vel}(c) and (d), from which we can see the air flow gathers at the surface center of the espresso cup and then flows upward with increasing speed. A similar phenomenon can be observed in our 2D simulation data in SM. 
Note that the velocity field is varying spatially and temporally. The maximum velocity ($v$-component) is approximately 0.4 m/s, while the mean value over the whole space-time domain is about 0.063 m/s.
To further examine the velocity and pressure inferred by PINN, we also evaluate the residuals of the governing equations, given in Eq.~(\ref{eq:PINN_residuals_3D}). The residuals of momentum equations are illustrated in Figure \ref{fig:Tomo_residual}. The values above the surface of the espresso cup are relatively higher than those away from the surface. However, the residuals are overall very small and the average over the spatial domain 
is in the order of $10^{-4}$~m/s$^{2}$, indicating that the velocity and pressure fields from Tomo-BOS/PINN satisfy the governing equations.

\subsubsection{An independent PIV experiment for validation}\label{sec:PIV_exp}

After obtaining the inferred velocity field, we also compare it with the diaplacement determined from schlieren-tracking. As shown in Figure \ref{fig:intro_2D_image}, we track distinct schlieren-features in sequential images, where the bottom of the structure moves from the second to the fifth frame by approximately 25~mm in $y$-direction, corresponding to a velocity magnitude of 0.08~m/s. This is much smaller than the inferred velocity of Tomo-BOS/PINN analysis. Therefore, we set up a PIV experiment to further examine the correctness of the PINN inference, especially the correctness of the maximal velocity. 

In PIV experiment, the espresso cup was placed within a glass box ($25\times25\times40$~cm$^3$) with an open top in order to provide a calm bath of air seeded with DEHS particles around the espresso cup.
Since the PIV experiment was performed independently to the Tomo-BOS experiment after a few months, the room temperature was 25$^{\circ}$C instead of 20$^{\circ}$C. A dual-head Nd:YAG laser (Litron NANO L 50-50, 532~nm) was used to generate a laser sheet of 10~cm in height with an average thickness of approximately 1~mm. A camera (LaVision Imager SX 6M, $2752\times2200$ pixels) equipped with a 50-mm lens acquired 400 double-frame images at 13~Hz with a $dt=1.5$~ms. The projected pixel resolution was 53~{$\mu m$/pixel}, resulting in a field of view of $13\times 10$~cm$^2$. Data was processed using an interrogation window size of $48\times48$ pixels with 75\% overlap, which resulted in $234\times189$ velocity vectors with a spacing of 640~$\mu$m.

\begin{figure}
\begin{center}
\includegraphics[width=0.95\textwidth]{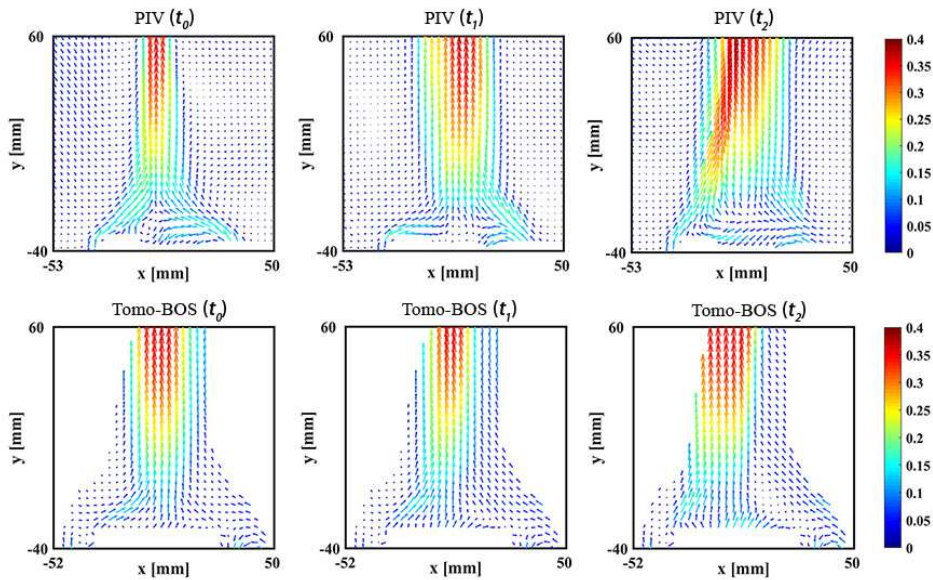}
\caption{Tomo-BOS/PINN and planar PIV results: 2D velocity vectors at various time instants. The 2D velocities at center plane $Z=-21$~mm are demonstrated for Tomo-BOS/PINN. The colormap represents the velocity magnitude of the vectors (unit: m/s). The results of Tomo-BOS/PINN shown here are at the instants $t=2.0$ s, $t=4.0$ s, $t=6.0$ s, while those of PIV are $t=1.0$ s, $t=7.0$ s, $t=16.0$ s. Movies
that demonstrate the time-dependent velocity fields are given in SM.
}
\label{fig:Tomo_PIV_BOS_vectors}
\end{center}
\end{figure}

\begin{figure}
\begin{center}
\includegraphics[width=0.7\textwidth]{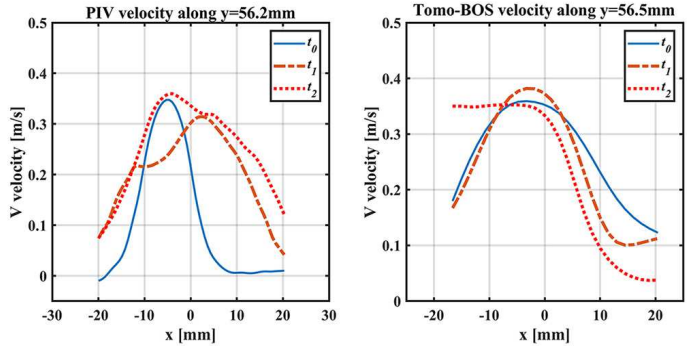}
\caption{Tomo-BOS/PINN and planar PIV results: velocity profiles along a horizontal line at various time instances. The profiles of Tomo-BOS/PINN results (right figure) are extracted from the center plane $Z=-21$~mm. 
}
\label{fig:Tomo_PIV_BOS_profiles}
\end{center}
\end{figure}

The velocity vectors in a 2D plane of PIV and Tomo-BOS/PINN experiments are demonstrated in Figure \ref{fig:Tomo_PIV_BOS_vectors}, where three typical time instants of each experiment are shown. Overall, we can observe similar flow pattern from two experiments and the velocity magnitudes are consistent with each other. The main difference between the PIV and the Tomo-BOS results is that there exist more variations right over the espresso cup in the PIV case.
In particular, there is a recirculation zone at $t_{0}$ in the PIV plots, which is not clear in the Tomo-BOS/PINN results. Moreover, there are some external disturbances from the right side in the PIV experiment, which can be seen at $t_{2}$ in Figure \ref{fig:Tomo_PIV_BOS_vectors}. 
As mentioned, the PIV and BOS experiments were performed independently and therefore a lining up between the PIV and Tomo-BOS/PINN results is not necessary. The comparison between the two experiments is meant to qualitatively validate the velocity by some  reference frames, thus the movements of the entire flow structure over the espresso cup are not relevant. 
The velocity profiles along a horizontal line at center plane at various time instances are also illustrated in Figure \ref{fig:Tomo_PIV_BOS_profiles}, in which we can observe the similarity in magnitudes and maximum velocities for all three profiles, and the similarity in width of the profiles between two experiments. The results also show that the unsteadiness of the flow does not affect the velocity magnitude in any of the experiments. 
In summary, the PIV experiment validates the Tomo-BOS/PINN in terms of the velocity range. Moreover, based on the results of Tomo-BOS/PINN and PIV, we can also conclude that tracking of the randomly visible schlieren structures in 2D images (shown in Figure \ref{fig:intro_2D_image}) cannot reliably determine flow velocities at specific locations due to its line-of-sight nature and consequently unknown location of the schlieren structure. 


\subsection{{Capability in dealing with sparse data}}\label{sec:TomoBOS_exp_sparse}

In Sect. \ref{sec:TomoBOS_exp_results}, PINN is applied to analyze the whole Tomo-BOS dataset which contains 400 snapshots with a recording rate of 50Hz. Here, we investigate the capability of PINNs in estimating high-resolution fields from sparse temperature data. 

\begin{figure}
\begin{center}
\includegraphics[width=0.7\textwidth]{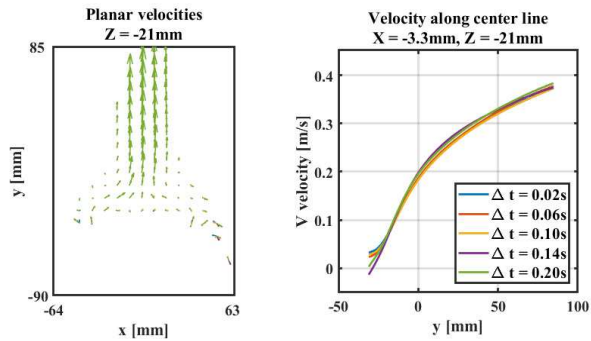}
\caption{Tomo-BOS/PINN result with temporal downsampling data: velocity fields at $t=2.0$~s inferred by PINN for various temporal resolutions. Left: 2D velocity vectors; right: velocity profiles of the $v$-component along the centerline. $\Delta t$ denotes the time interval between two consecutive snapshots used for PINN algorithm. 
}
\label{fig:3Dresults_vel_timestep}
\end{center}
\end{figure}

\begin{figure}
\begin{center}
\includegraphics[width=0.9\textwidth]{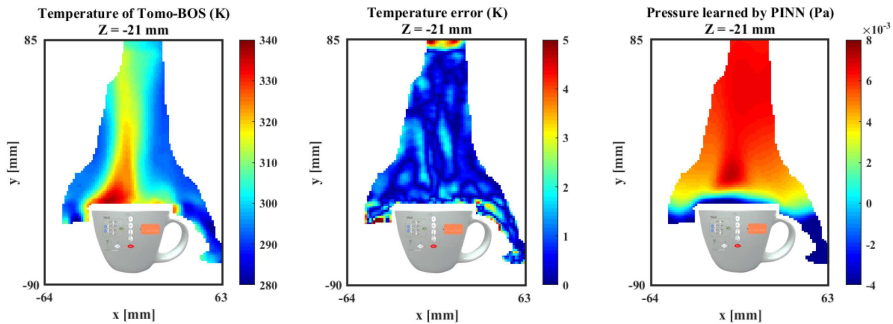}
\caption{Tomo-BOS/PINN result with temporal downsampling data: BOS temperature, absolute error of temperature, inferred pressure at $z=-21$ mm and $t=2.96$ s. The training data is sampled with time interval $t=0.1$~s. Although this snapshot is not used in training (unseen data for PINN), the relative $L_2$-norm error of temperature for this snapshot is 0.362\%. 
}
\label{fig:3D_testDT_2Dfields}
\end{center}
\end{figure}

We first perform temporal downsampling by extracting the temperature from Tomo-BOS data with different time intervals, from which we infer the velocity and pressures fields. As shown in Figure \ref{fig:3Dresults_vel_timestep}, the velocity fields and profiles from different datasets are consistent, indicating that the proposed algorithm is not sensitive to the data sparsity. We note $\Delta t=0.2$s 
means that a maximum displacement between two consecutive snapshots is about 0.08~m, which is even larger than the characteristic length $L = 0.065$~m. 
Moreover, as we leave out some temperature frames during training, we can use them for validation. Specifically, if the training data is sampled with $\Delta t=0.1$~s, we can compute the temperature error for those intermediate frames, as demonstrated in Figure \ref{fig:3D_testDT_2Dfields}, which shows the 2D temperature, pressure and velocity fields at $t=2.96$ s. Although this snapshot is not used in training (unseen data for PINN), the relative $L_2$-norm error of 3D temperature for this testing frame is only 0.362\%.

\begin{figure}
\begin{center}
\includegraphics[width=0.9\textwidth]{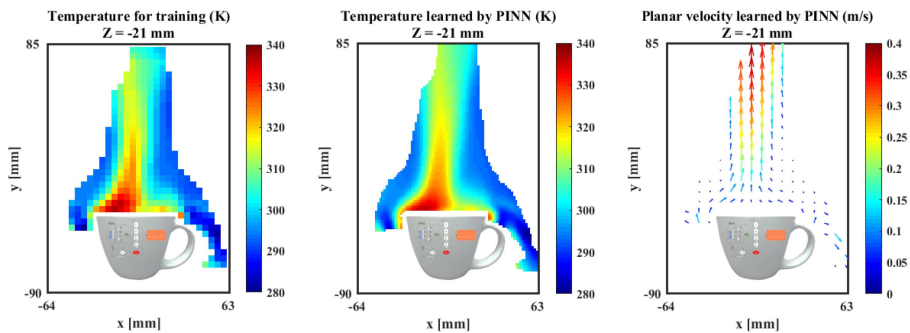}
\caption{Tomo-BOS/PINN result with spatial downsampling data: (from left to right) downsampled temperature data used for PINN, temperature regressed by PINN and velocity vectors inferred by PINN at $z=-21$~mm and $t=2.96$~s. The training data is downsampled along each direction, which results in a grid resolution of 3.0~mm, while that of the reconstructed fields is 1.5~mm. Although the data for PINN at each snapshot only contains 1/8 of the Tomo-BOS data, the relative $L_2$-norm errors of temperature for this snapshot is 0.430\%. 
}
\label{fig:3D_space30_2Dfields}    
\end{center}
\end{figure}

Furthermore, we can also downsample the temperature data in space, as illustrated in Figure \ref{fig:3D_space30_2Dfields}. 
In particular, the temperature data is downsampled along each direction with a grid resolution of 3.0~mm, while that of the original Tomo-BOS data is 1.5~mm. By doing so, the training data at each frame only contains 1/8 of the Tomo-BOS data. 
We find from Figure \ref{fig:3D_space30_2Dfields} that PINN is able to infer the full temperature and velocity fields with good consistency. 
The relative $L_2$-norm errors of temperature for the frame is 0.430\%. 
The assessments in this section indicate that the PINN algorithm allows us to reconstruct continuous and high-resolution velocity and pressure fields from relatively sparse data (temporally and spatially) due to the introduction of the governing equations. 
In addition, one can take advantage of downsampling since it can help to improve the efficiency of the training process without reducing the accuracy significantly.
A systematic study to quantify the accuracy of PINNs against data resolution is performed on the 2D simulated case, which can be found in SM. 

\subsection{Discussion on the physical properties}\label{sec:TomoBOS_exp_discussion}

\begin{figure}
\begin{center}
\includegraphics[width=\textwidth]{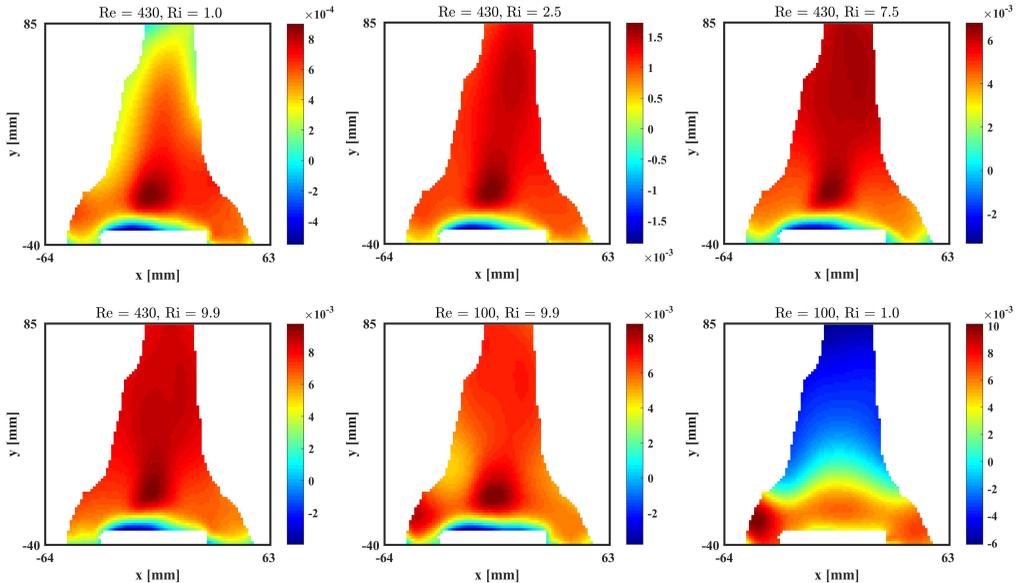}
\caption{Sensitivity of physical properties used in PINN: 2D pressure fields ($p-\bar{p}$) inferred by using various parameters at $z=-21$~mm and at $t=2.0$~s. 
}
\label{fig:Tomo_pressure_parameters}
\end{center}
\end{figure}

\begin{figure}
\begin{center}
\includegraphics[width=\textwidth]{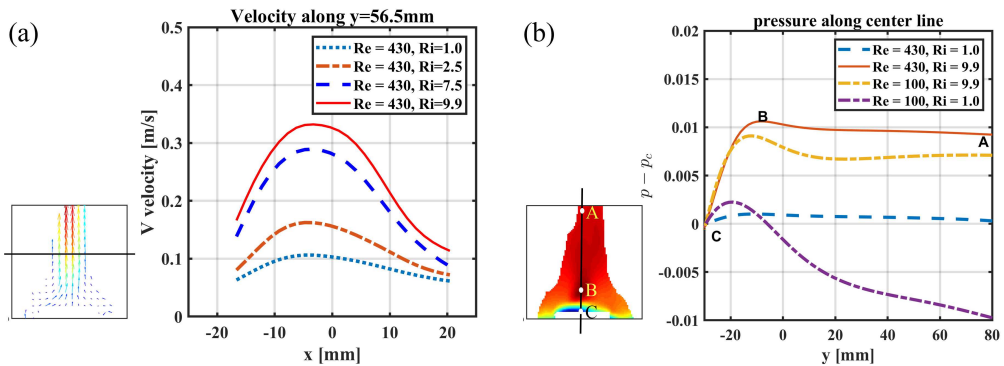}
\caption{Sensitivity of physical properties used in PINN: (a) velocity profiles ($v$-component) along a horizontal line with various Richardson number, (b) pressure profiles ($p-p_{c}$) along the center line with various parameters. The profiles are extracted from the plane $z=-21$~mm and time instance $t=2.0$~s. Note that the training data for different conditions are the same. 
}
\label{fig:Tomo_profiles_parameters}
\end{center}
\end{figure}

The non-dimensional parameters defined in Eq.~(\ref{eq:non_dimensional_para}) are $\mbox{Re} = 430$, $\mbox{Pe} = 313$, $\mbox{Ri} = 9.9$, which are derived from the physical properties given in Sect. \ref{sec:TomoBOS_PINN}. 
As the underlying physical laws are encoded in the neural network, the estimation result is generally influenced by these non-dimensional parameters. In this section, we investigate the sensitivity of Tomo-BOS/PINN to the values of Re and Ri ($\mbox{Pe}=0.73\mbox{Re}$).
We note that these parameters, which are sensitive to the physical properties (e.g., viscosity, conductivity) as well as the reference temperatures, can be affected by one or more than one factors. Instead of tuning these factors one by one, we select the non-dimensional parameters from a proper range to cover different conditions. Specifically, we consider the Reynolds number $\mbox{Re}=\{100, 430\}$ and the Richardson number $\mbox{Ri}=\{1.0,2.5,7.5,9.9\}$. 
The pressure fields of
different combinations are shown in Figure \ref{fig:Tomo_pressure_parameters}. 
Increasing Ri, the pressure over the espresso cup also increases, with the highest pressure generally located over the cup, but not right on the surface. In addition, larger Ri leads to larger velocity magnitude, which can be observed in Figure \ref{fig:Tomo_profiles_parameters}(a).
Somewhat different is the case with $\mbox{Re} = 100$ and $\mbox{Ri} = 1$, where we find an inverse pattern of the pressure field, as shown in Figure \ref{fig:Tomo_profiles_parameters}(b), which shows the pressure profiles along the center line for various parameters. 
Note that the velocity and pressure fields are inferred by integrateing the temperature data and the governing equations, where the temperature data used for different non-dimensional configurations are the same. We conclude that the inference results of PINN are relatively sensitive to the parameters used in the encoded governing equations, especially to the Richardson number for the investigated buoyancy-driven flow. 
For future development of the proposed method, a Bayesian framework of PINN~\citep{yang2021b} can be considered to address the issue caused by model uncertainty.

\section{Concluding Remarks} \label{sec:conclusion}

In this paper, we propose a machine learning algorithm based on physics-informed neural networks for estimating velocity and pressure fileds from temperature data of Tomo-BOS experiments. 
The advantages of PINN algorithm are summarized as follows. (a) PINNs are capable of integrating the governing equations and the temperature data, which is similar to the variational data assimilation methods, while it is not necessary to solve the governing equations by using any CFD solvers. (b) PINNs can infer the velocity and pressure simultaneously by regressing the data, without any information of the initial and boundary conditions. (c) PINNs can provide continuous solutions of the velocity and pressure, even if the experimental data are sparse and limited.

We first evaluate the proposed method by using a 2D synthetic simulation of buoyancy-driven flow, which can be found in the supplementary material, where the temperature is generated by CFD solver directly. The performance of PINNs against various parameter settings (including the size of neural network, the weighting coefficient in the loss function, the spatial and temporal resolutions and the data noise level) is investigated systematically. This parametric study, which helps us understand how to use the algorithm in a good manner, reveals that the method can provide accurate and reliable performance on extracting flow fields from temperature against sparse data. Moreover, the proposed PINN algorithm can also handle noisy data to some extent. It is also possible to extend the proposed method to a Bayesian-PINN framework~\citep{yang2021b}, which can address large noise level and quantify the uncertainty of the inference result simultaneously. 

We then present a Tomo-BOS experiment to observe the flow over an espresso cup. The 3D velocity and pressure fields are successfully inferred from the reconstructed 3D temperature data, as qualitatively validated by a comparison with the velocity obtained from PIV measurements. 
Compared to PIV, the BOS technology is capable of straightforwardly providing enormous quantities of data on the investigated fluid flows, which has become increasingly popular due to its simplicity. 
Here, we demonstrate that either planar or tomographic BOS data is simply available to PINN algorithm for velocity and pressure quantification, which is novel and can open a door to BOS velocimetry for complex fluid flows. 
Furthermore, the flexibility of the proposed method - PINNs - allows us to extend the algorithm to different types of flows (in addition to the natural convection investigated in this paper) by simply encoding proper governing equations of the fluid flows. 
Taken together, the results in this paper indicate that the proposed method is accurate and flexible in dealing with data of various fluid mechanics problems, which can become a promising direction in experimental fluid mechanics.

\section*{Declaration of interests}
The authors report no conflict of interest

\section*{Acknowledgements}
The authors S. Cai, Z. Wang and G.E. Karniadakis acknowledge the support from the PhILMS grant under the grant number DE-SC0019453.

\bibliographystyle{jfm}
\bibliography{jfm-instructions}


\end{document}